\documentclass[12pt]{article}
\usepackage{epsfig}
\usepackage{graphicx}

\newcommand\pubnumber{SLAC-PUB-11857}
\newcommand\pubdate{May, 2006}
\newcommand\hepnumber{hep-ph/0605179}

\def\SLAC{Stanford Linear Accelerator Center\\
  Stanford University, Stanford, California 94309 USA}
\def\doeack{\footnote{Work supported by the US Department of Energy,
                     contract DE--AC02--76SF00515.}}

\def\Title#1{\begin{center} {\Large #1 } \end{center}}
\def\Author#1{\begin{center}{ \sc #1} \end{center}}
\def\Address#1{\begin{center}{ \it #1} \end{center}}

\newcommand\pubblock{\rightline{\begin{tabular}{l} \pubnumber\\
         \pubdate \\ \hepnumber \end{tabular}}}

\begin{document}
  \begin{titlepage}
  \pubblock
  
  \vfill
  \Title{Finite Temperature Corrections to Relic Density Calculations}
  \vfill
  \Author{Tommer Wizansky\doeack}
  \Address{\SLAC}
  \vfill
  \begin{abstract}
    In this paper we evaluate finite temperature corrections to the dark matter
    relic density within the context of minimal supersymmetry with a neutralino LSP. We 
    identify several regions of parameter space where the WIMP annihilation cross section is
    especially sensitive to small corrections to the undelrlying parameters. 
    In these regions, finite temperature effects have the potential to be important. 
    However, we shall show by explicit calculation that these effects are small. 
    In the regions we investigated, the maximal corrections are on the order
    of $10^{-4}$ and are therefore negligible compared with theoretical and experimental 
    uncertainties.
  \end{abstract}
  \vfill
  \end{titlepage}
  \def\thefootnote{\fnsymbol{footnote}}
  \setcounter{footnote}{0}

  \section{Introduction}
  
  It is now well accepted that approximately 20\% of the matter in the universe
  is in the form of a new non-baryonic neutral and weakly interacting particle 
  whose microscopic properties have yet to be determined. 
  This material has been termed dark matter. 
  Experimentally, the ratio of the dark matter density, on cosmological scales, to the critical 
  denstiy has been measured within 10\% to be $\Omega_{DM}\sim0.2$ \cite{Spergel}.  
  Given a theory of dark matter, the relic density can be predicted by caculating the particle's
  annihilation cross sections and integrating the Boltzmann equation through decoupling to 
  the present day.  It is interesting to note that if we assume that the dark matter
  particle  annihilates through weak interactions and has a mass on the order of the 
  electroweak scale, we obtain the correct value for the relic density 
  (see, for example, \cite{Kolb:1990}).  This fact has led many theorists to concentrate 
  on weakly interacting massive particle (WIMP) models of dark matter.  

  Assuming that dark matter is composed of WIMPs that are in thermal equilibrium in the
  early universe, their density today is given approximately by the expression
  \cite{Scherrer:1985zt}
  \begin{equation}
    \Omega = \frac{s_0}{\rho_c}\left(\frac{45}{\pi g_*}\right)^{1/2}\frac{x_f}{m_{pl}}
    \frac{1}{\langle\sigma v\rangle}
  \end{equation} 
  where $s_0$ is the current entropy density of the universe, $\rho_c$ is the critical density,
  $g_*\approx80$ is the number of relativistic degrees of freedom at the time of freeze-out, $\langle
  \sigma v\rangle$ is the thermally averaged annihilation cross section and $x_f=m/T$ is the
  ratio of the WIMP mass to the freeze-out temperature. For weak scale interactions 
  $x_f\sim25$. From the above value of $x_f$ it is clear that the WIMP is non relativistic
  at the time of freeze-out and for this reason it has long been supposed that finite temperature
  effects can be neglected when calculating the annihilation cross section.  
  Indeed, these corrections tend to be suppressed either exponentially, or by several powers of $T/m$ 
  (see, for example, \cite{Matsumoto:1999kd}). 
  However, currently favored models of dark matter can often lie in regions of
  parameter space where the annihilation cross section is extremely sensitive to small variations
  in the underlying parameters. For this reason we consider the issue of  finite temperature 
  corrections to the relic density worth revisting.

  For definiteness we consider models of dark matter within the context of the minimal 
  supersymmetric standard model (MSSM). Dark matter scenarios in the MSSM have been 
  discussed by many authors.  In paricular, the parameter space of a subclass of models -
  minimal supergravity (mSUGRA) - has been extensively mapped out 
  \cite{Ellis:2000we,Edsjo:2003us,Baer:2003ru}. Viable mSUGRA models fall roughly into
  four parameter regions: the bulk region, the focus point region, the coannihilation 
  region and the A-funnel. In the bulk region the superpartners are all
  relatively light and the annihilation is dominated by lepton production through P-wave t-channel 
  and s-channel slepton exchange.  In the focus point region the scalar superpartners are 
  very heavy. In this case the LSP obtains a significant
  higsino and wino component and the annihilation is dominated by s-channel W and Z production.  
  The coannihilation region is
  characterized by a nearly degenerate next-to-lightest super partner (NLSP).  In this scenario 
  the NLSP has a finite density during freeze-out and coannihilates with the LSP. Finally, the 
  A-funnel region is characterized by $m_A\sim2m_{N_1}$ where $m_A$ is the mass of the CP-odd heavy Higgs
  $A^0$ and $m_{N_1}$ is the mass of the LSP. In this region of parameter space the WIMP 
  annihilations proceed mainly through resonant s-channel $A^0$ exchange. 
  
  Three of the above regions -- the bulk, the coannihilation region and the A funnel -- exhibit 
  sensitivity to one or more of the underlying MSSM parameters.  In the bulk, the S-wave 
  annihilation is helicity suppressed. The dominant channel is thus P-wave, 
  which is itself suppressed by two powers
  of the WIMP velocity, $v$. Any enhancememt to the helicity breaking masses of the standard 
  model fermions would thus significantly enhance the overall annihilation 
  cross section. In the A-funnel the WIMP 
  annihilation cross section is naturally highly sensitive to the width of the $A^0$ boson. 
  Finally, in the 
  coannihilation region the cross section is sensitive to the abundance of the NLSP, which is 
  exponentially dependent on it mass. In all these cases small 
  corrections due to finite temperature effects have the potential to be significant.

  Similar issues, in which dark matter annihilation cross sections are tuned by the 
  adjustment of mass differences or particle mixings occur in many other models, both
  within and outside of the context of supersymmetry. 

  In this paper we will present specific estimates of the finite temperature effects on 
  the relic denstiy in the scenarios that we have discussed. The outline of the paper is 
  as follows: In Section \ref{section:ThermalFieldTheory} we review elements
  of finite temperature field theory relevant to our analysis.  In Section \ref{section:MassSplitting} 
  we calculate the thermal corrections to the abundance of the NLSP at freeze-out in the stau
  coannihilation region. In Section \ref{section:FermionMasses} we analyze the enhancements to the 
  helicity breaking standard model fermion masses and their effect on the relic density.  In Section 
  \ref{section:AWidth} we evaluate the thermal corrections to the width of the $A^0$ boson in the 
  A-funnel region. Finally, in Section \ref{section:conclusions} we review our analysis and present
  conclusions.

  Other corrections to relic density predictions stemming from a better understanding of the high
  temperture environment in the early universe have been suggested. Most recently, Hindmarsh 
  and Philipsen \cite{Hindmarsh:2005ix} have analyzed the deviation of the QCD 
  equation of state from that of an
  ideal gas. They found that this effect could lead to measurable a correction to the dark 
  matter abundance.  It is, however, beyond the scope of this paper to discuss these recent
  developments.

  \section{Thermal Field Theory}
  \label{section:ThermalFieldTheory}
  At finite temperatures, field theory calculations must reflect the fact that all
  correlation functions are thermally averaged. The expression 
  $$
    \langle0|\mathcal{O}|0\rangle
  $$
  for the zero temperature vaccum expectation value of an operator must be replaced by
  $$
    \frac{\sum_n\langle n|\mathcal{O}e^{-\beta H}|n\rangle}{\sum_n\langle n|e^{-\beta H}|n\rangle}.\\
  $$
  where the sum is over a complete set of states. Perturbation theory then leads to a set of 
  corrected Feynman rules for S-matrix elements. To a good level of accuracy the corrections to 
  relic density calculations can be understood through the thermal shift in physical 
  parameters such as
  the masses and widths of interacting particles.  For this reason we will concentrate on the 
  corrections to the real and imaginary parts of propagators and we do not find it necessary 
  to evaluate general finite temperature Green's functions. 

  Two parallel formalisms exist for perturbative calculations at finite temperatures; 
  the imaginary time formalism (ITF) which was first introduced by Matsubara 
  \cite{Matsubara:1955ws} incorporates a contour along the imaginary
  time axis to obtain the thermal average. This technique is best suited for the calculation of 
  equilibrium properties of hot plasmas but can be extended through analytic continuation to real time 
  correlation functions. The real time formalism (RTF) was developed later by Schwinger 
  \cite{Schwinger:1960qe}, Mills \cite{Mills:1961} and Keldysh \cite{Keldysh:1964ud}
  and incorporates dynamics more naturally. The latter also provides more 
  calculational ease. It is this technique which we shall utilize in the following.
  
  In the RTF the degrees of freedom in a theory are doubled.  The new fields do not create 
  or annihilate 
  external particles but may appear in loops. The new fields do not couple to the
  physical fields through Feynman diagram vertices but rather mix with them through 
  propagators.  It follows that these new fields do not appear at one loop level, and we can ignore 
  them in this analysis. Then, the propagators of scalars, chiral 
  fermions, and gauge bosons, respectively are \cite{Kobes:1984vb} 
  \begin{eqnarray}
    \label{eqscalarprop}
    D(P) & = & \frac{i}{P^2-m^2+i\eta}+2\pi\delta(P^2-m^2)n(p_0) \\
    \label{eqfermionprop}
    S(P) & = & (\sigma\cdot P)\left[\frac{i}{P^2-m^2+i\eta}-
      2\pi\delta(P^2-m^2)\tilde{n}(p_0)\right] \\
    \label{eqgaugeprop}
    D_{\mu\nu}(P) & = & -g_{\mu\nu}\left[\frac{i}{P^2-m^2+i\eta}+
      2\pi\delta(P^2-m^2)n(p_0)\right].
  \end{eqnarray}
  We have used Feynman gauge and the fermionic propagator is given in two component 
  notation. The above equations contain the Bose-Einstein and Fermi-Dirac distributions, 
  respectively,
  \begin{eqnarray*}
    n(p_0) & = & \frac{1}{e^{\beta |p_0|}-1} \\
    \tilde{n}(p_0) & = & \frac{1}{e^{\beta |p_0|}+1}.
  \end{eqnarray*}
  Throughout this paper, upper case letters will denote four momenta and lower case letters
  will be reserved for the spatial components.
  If the fermionic Lagrangian contains a helicity breaking mass term we will obtain a 
  corresponding Feynman rule for a mass insertion, given by
  \begin{eqnarray}
    \langle\psi_L\psi_R^{\dagger}\rangle & = & -m^*c\left[\frac{i}{P^2+i\eta}-
      2\pi\delta(P^2-m^2)\tilde{n}(p_0)\right]
  \end{eqnarray}
  where c is the antisymmetric $2\times2$ matrix defined so that $c_{12}=-1$.
  To evaluate the imaginary part of self energy diagrams we will use a generalization of the zero 
  temperature cutting rules, due to Kobes and Semenoff \cite{Kobes:1986za}. As in the zero
  temperature case, the
  imaginary part of a diagram is obtained by replacing full propagators by ``cut'' propagators.
  For fermions these are given by
  \begin{eqnarray}
    \label{eqnarraycutprops1}
    S^{>}(P) & = & 2\pi\epsilon(p_0)(1-\tilde{f}(p_0))\delta(P^2-m^2) \\
    \label{eqnarraycutprops2}
    S^{<}(P) & = & -2\pi\epsilon(p_0)\tilde{f}(p_0)\delta(P^2-m^2)
  \end{eqnarray}
  where $\epsilon$ is the sign function and
  $$
  \tilde{f}(p_0) = \frac{1}{e^{\beta p_0}+1}.
  $$ 
  Note the absence of the absolute value in the 
  argument of the exponent.

  Throughout this paper, we will take the plasma to be comprised solely of light 
  ($m_f\leq10$ GeV)
  standard model fermions and massless gauge bosons. We will consistently neglect terms
  representing interactions with superparticles, heavy gauge bosons and heavy Higgs
  bosons in the thermal plasma. This is a reasonable approximation since such 
  terms will be exponentially suppressed by the ratio of their mass to the 
  freeze-out temperature.

  \section{The LSP-NLSP Mass Splitting}
  \label{section:MassSplitting}
  An often considered scenario for coannihilation involves the stau 
  as the NLSP. In this section we will analyze the finite temperature corrections
  to the model. The annihilation rate is exponentially dependent on 
  the mass splitting between the stau and the lightest superpartner. 
  Thus it is interesting to compute the thermal corrections to 
  this mass splitting. 

\begin{figure}[t]
    \begin{center}
      \leavevmode
      \epsffile{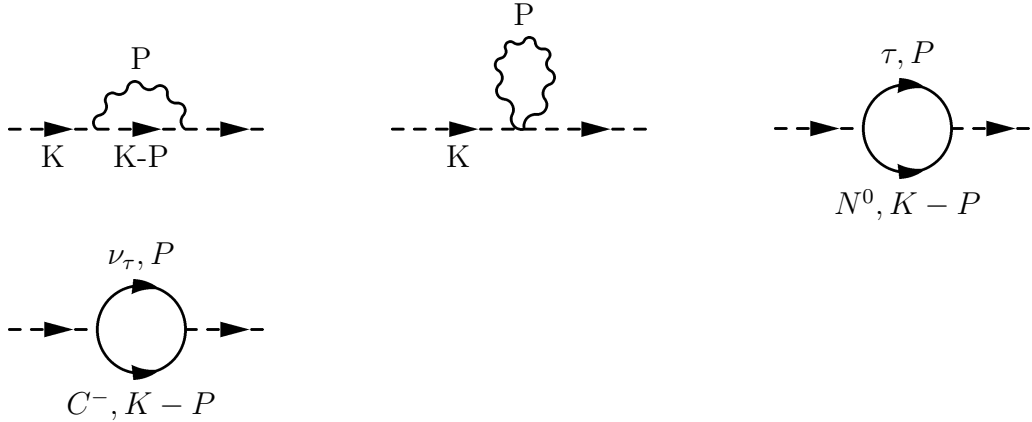}
    \end{center}
    \caption{One loop diagrams contributing to the self energy of the stau.}
    \label{figstauselfenergy}
  \end {figure}

  The full sfermion propagator is given by
  $$
  \frac{-i}{K^2-m_0^2-\Pi(K)^2}.
  $$
  Where $\Pi(K)$ is the one paricle irreducible self energy. The particle's dispersion relation, 
  $k_0(k)$ is obtained by solving the equation  
  $$
  k_0^2-k^2-m_0^2-\Pi(K)^2=0
  $$
  and the mass shift is then given by the limit $\delta m=k_0-m_0\Big|_{k=0}$. 
  At one loop level, the thermal corrections to the stau mass originate from the Feynman 
  diagrams depicted in Fig.~\ref{figstauselfenergy}. Each propagator has two terms, shown in 
  eqs. (\ref{eqscalarprop})-(\ref{eqgaugeprop}).  The first represents a contribution from a 
  virtual particle
  and the second from a thermal particle. The first term is the zero-temperature component of the
  propagator. 
  To obtain the thermal contributions to the real part of a one loop diagram we need to consider terms
  involving one thermal and one virtual field. As mentioned, terms involving thermal heavy particles 
  will be suppresses by small statistical factors and we can safely neglect them.

  We begin with the photon loop.  Using (\ref{eqscalarprop}) and (\ref{eqgaugeprop}),
  \begin{equation}
    -i\Pi^{\gamma-loop}=(ie)^{2}\int\frac{d^{4}P}{(2\pi)^4}(K-P+K)^{\mu}
    D_{\mu\nu}(P)D(P-K)(K-P+K)^{\nu}
  \end{equation}
  We write the real part of the self energy and ignore the $T=0$ terms as well 
  as those representing interactions with sfermions in the plasma. This gives
  \begin{equation}
    Re\Pi^{\gamma-loop}_\beta=ie^{2}Q^{2}\int\frac{d^{4}P}{(2\pi)^4}(4K^2+P^2-4K\cdot P)
    \left[\frac{i2\pi\delta(P^2)}{((P-K)^2-m_0^2)}n(p_0)\right]
  \end{equation}
  where $m_0$ is the zero temperature mass of the stau and the subscript $\beta$ denotes 
  the thermal part of the diagram. This can be evaluated to give 
  \begin{equation}
    Re\Pi^{\gamma-loop}_\beta \quad = \quad -\frac{\alpha}{2\pi}\int dp\,p\left[
      2+(L_1^{+}(m_0)+L_1^{-}(m_0))\frac{K^2+m_0^2}{4kp}
      \right]n(p)
  \end{equation}
  where, following \cite{Levinson:1985ub}, we have defined
  \begin{equation}
    \label{eqL1}
    L_1^\pm(K,p,m) \equiv\pm\ln\left(\frac{K^2-m^2\pm 2(k_0+k)p}{K^2-m^2\pm 2(k_0-k)p}
    \right).
  \end{equation}
  Note that
  \begin{equation}
    \label{eqlimL}
    \mathop {\lim }\limits_{k \to 0} \frac{L_1^\pm}{k}  = \frac{4p}{k_0^2\pm 2k_{0}p-m_0^2}. 
  \end{equation}
  To first order in $\alpha$ we can set $k_0(0)=m_0$ inside the integral, which then 
  considerably simplifies. We obtain for the mass shift
  \begin{equation}
    \label{eqdmstau1}
    \delta(m^2)^{\gamma-loop} = -\frac{\alpha}{\pi}\int dp\,p\,n(p) = -\frac{\pi\alpha}{6}T^2.
  \end{equation}
  
  The expression for the photon tadpole is
  \begin{equation}
    -i\Pi^{tadpole}=ig^2\int\frac{d^{4}P}{(2\pi)^4}D_{\mu}^{\mu}(P).
  \end{equation}
  Evaluating this and taking the limit as $k\to0$ we obtain
  \begin{equation}
    \delta(m^2)^{tadpole}=\frac{4\pi\alpha}{3}T^2.
  \end{equation}

  Next, we compute the lepton-gaugino loops.  Using (\ref{eqfermionprop}) and following the same steps
  as those for the photon loops we obtain for a digaram containing virtual neutralinos
  \begin{equation}
    \label{eqdmstau2}
    \delta(m^2)^{tau-N} = \frac{8\alpha}{\pi} m_{\tilde{\tau}}^2\sum_{i}|G_{\tau Ni}|^2
    \int dx\frac{x^2y}{(1-\mu_i^2)^2-4y^2}\frac{1}{e^{y/t}+1}
  \end{equation}
  where $G_{\tau N_i}$ are the couplings of the tau stau and neutralino, 
  $\mu_i=m_i/m_{\tilde{\tau}}$ 
  is the mass of the $i^{th}$ neutralino scaled by the stau mass, 
  $\mu_\tau=m_\tau/m_{\tilde{\tau}}$, $y=\sqrt{x^2+\mu_\tau^2}$, $t=T/m_{\tilde{\tau}}$ 
  and the sum runs over the four neutralino types.  
  In the derivation of this expression we have consistently neglected terms that
  are proportional to the neutralino statistical factor. A similar expression gives 
  the contribution from the diagrams with virtual charginos.
  
\begin{figure}[t]
    \begin{center}
      \leavevmode
      \epsffile{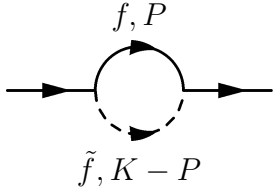}
    \end{center}
    \caption{Feynman diagram contributing to the neutralino self energy at one loop order
      and in the limit of a plasma populated by light SM particles.}
    \label{figNselfenergy}
  \end{figure}

  In the limit of the plasma being populated solely by light SM particles, the neutralino
  obtains a thermal mass shift from self energy diagrams of the type depicted in 
  Fig.~\ref{figNselfenergy}, with a virtual sfermion and a thermal standard model
  fermion. Note that we have neglected helicity flipping diagrams since these will be
  suppressed by light fermion masses.
  
  \begin{equation}
    \label{eqsNmassfloop}
    -i\Sigma = -(ie)^2\sum_{f}|G_{fN}|^2\int\frac{d^4P}{(2\pi)^2}cS(P)cD(K+P)
  \end{equation}
  where the sum runs over all fermion species inhabiting the plasma and $G_{fN}$ is 
  the neutralino-fermion-sfermion coupling.

  In order to extract the mass shift we should better understand the structure of 
  this expression. The most general form of the 4-component self 
  energy allowed by the Majorana nature of the neutralino is
  \begin{equation}
    \Sigma_4(K) = aK_\mu\gamma^\mu+bu_\mu\gamma^\mu\gamma^5+c.
  \end{equation}
  where $u_\mu$ is the 4-velocity of the thermal bath. In the approximation that 
  we ignore the helicity flip diagrams, $c=0$. Then, the dispersion relation,
  $$
    \det(K\cdot\gamma - \Sigma_4(K) - m_0)=0
  $$
  taken at zero momentum, leads to a thermal mass given by the equation
  \begin{equation}
    (1-a)^2m^2=(m_0)^2+b^2.
  \end{equation}
  Examining (\ref{eqsNmassfloop}) we see the $\Sigma(K)=-\Sigma(-K)$ so we can conclude
  that $b=0$. The identity
  $$
  \delta m=am=\frac{1}{2}Tr\left\{\mbox{Re}\,\Sigma\right\}.
  $$ 
  then follows.  Evaluating this, we obtain
  \begin{equation}
    \label{eqdmN}
    \delta m = -\frac{2\alpha m_N}{\pi}|G_{fN}|^2\int dx\,\frac{x^2y}
    {(1-\mu_{\tilde{f}}^2)^2-4y^2}\frac{1}{e^{y/t}+1}.
  \end{equation}
  where here $y=\sqrt{x^2+\mu_f^2}$.
  
\begin{figure}[t]
    \begin{center}
    \leavevmode
    \epsffile{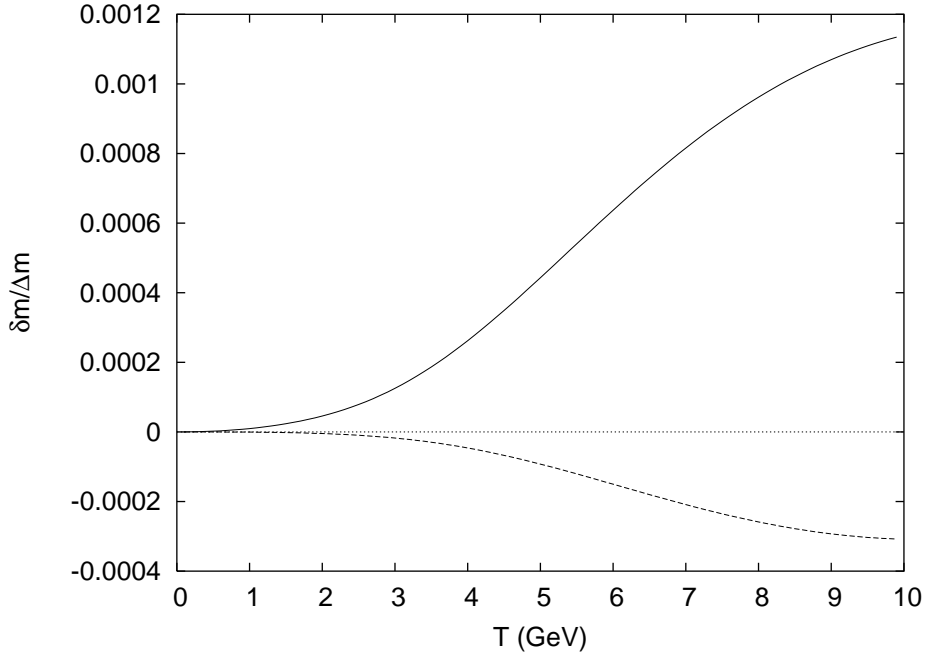}
    \end{center}
    \caption{Thermal corrections to the masses of the stau (solid) and lightest
      neutralino (dashed) in the stau coannihilation region.  The values are scaled
      by the zero-temperature mass splitting.}
    \label{figstaumassshift}
   \end{figure}

  Using (\ref{eqdmstau1}), (\ref{eqdmstau2}) and (\ref{eqdmN}) We can calculate the 
  correction to the stau-neutralino 
  mass splitting.  A typical, well studied, point in the coannihilation region is given by benchmark
  point LCC3 which is described in detail in \cite{Khotilovich:2005gb}. 
  At this point the mass splitting
  between the stau and the lightest neutralino is approximately 10 GeV. In 
  Fig.~\ref{figstaumassshift}  we plot the total shift in this mass splitting, at LCC3, as a 
  function of temperature. At freeze-out, which occurs at approximately 5 GeV, thermal 
  effects account for a fractional correction of about $10^{-4}$. The WIMP annihilation
  rate is determined by the relative Boltzman factor between the two particles, which
  is then corrected by
  \begin{equation}
    e^{-(m_{\tilde\tau}-m_N)/T}\rightarrow e^{-(m_{\tilde\tau}-m_N)/T}
    e^{-(m_{\tilde\tau}-m_N)10^{-4}/T}
    \approx e^{-(m_{\tilde\tau}-m_N)/T}(1-2\times10^{-4}).
  \end{equation}
  We evaluated the effect of the shift in mass splitting on the relic density using
  the DarkSUSY \cite{Gondolo:2004sc} package.  We found, as expected, that 
  the change in the relic density prediction is roughly
  \begin{equation}
    \frac{\delta\Omega_{DM}}{\Omega_{DM}}\approx1.0\times\frac{\delta(\Delta m)}{\Delta m}.
  \end{equation}

  \section{Helicity Breaking Fermion Masses}
  \label{section:FermionMasses}
  It would seem that a straightforward way for two WIMPs to annihilate would be by production of fermion pairs
  through t-channel and u-channel sfermion exchange.  For gravity mediated supersymmetry breaking
  this is especially true since the LSP is typically almost pure bino and as such does not
  couple to the Higgs boson or to gauge bosons.  But S-wave annihilation of neutralinos into
  fermion-antifermion pairs is helicity suppressed by the square of the 
  ratio of the mass of the fermion to the mass of the LSP \cite{Goldberg:1983nd}.  
  In the bulk region of the mSUGRA
  parameter space, this number is on the order of $10^{-4}$ for tau or bottom quark 
  production. P-wave annihilation is not helicity suppressed but is proportional 
  to two powers of the WIMP velocity.  This is typically about $(T/m_N)\approx10^{-2}$. 
  Thus, the helicity suppressed S-wave processes contribute to the annihilation at a percent
  level. 

  Thermal effects cannot on their own give fermions helicity breaking masses 
  but they do have the 
  potential to enhance (or suppress) existing mass operators. The thermal corrections to these
  operators must vanish as the zero temperature mass becomes small and will thus be
  proportional to it. It is possible that the coefficient of proportionality could be large since 
  the SM fermions are light compared to the ambient temperature.
  
\begin{figure}[t]
    \begin{center}
      \leavevmode
      \epsffile{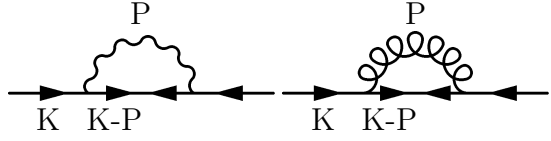}
    \end{center}
    \caption{Feynman diagrams contributing, at one loop order, to the helicity breaking 
      mass of the standard model fermions.}
    \label{fighbmassshiftSMf}
  \end{figure}

\begin{figure}[!t]
    \begin{center}
    \leavevmode
    \epsffile{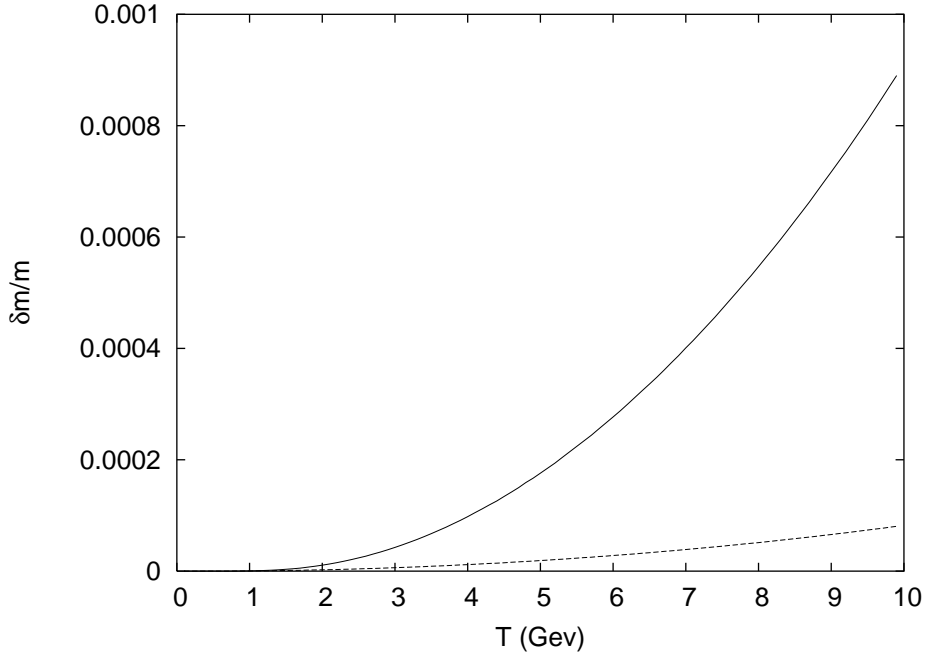}
    \end{center}
    \caption{Thermal correction to the helicity breaking mass operator of the bottom quark
      (solid) and tau lepton (dashed).}
    \label{figbhflipshift}
  \end{figure}

  Assuming that only the light gauge bosons and SM fermions are thermal, the diagrams contributing 
  to the fermion helicity breaking masses are those depicted in Fig.~\ref{fighbmassshiftSMf}. 
  For the photon loop

  \begin{equation}
    -i\Sigma = (ie)^2\int \frac{d^4P}{(2\pi)^4} \bar\sigma^\mu 
    S(K-P)D(P)_{\mu\nu}\bar\sigma^\nu.
  \end{equation}
  We evaluate the integral and take the real part of $\Sigma$, neglecting the zero temperature 
  terms, to obtain
  \begin{equation}
    \label{eqhfphoton}
    Re\Sigma_{\beta} = \frac{\alpha}{\pi^3}m_fc\int d^4P\left[
      \frac{2\pi\delta(P^2)}{(K+P)^2-m_f^2}-
      \frac{2\pi\delta(P^2-m_f^2)}{(K+P)^2}\right]n(p_0)
  \end{equation}
  where $m_f$ is the fermion mass. We can perform the $p_0$ integration using the delta function.  
  The first term in brackets yields
  \begin{eqnarray*}
    Re\Sigma_{\beta}^{(1)} & = & \frac{\alpha}{\pi^2}m_fc\int\frac{d^3p}{p}\left[
      \frac{1}{K^2+2(k_0-k\cos\theta)p-m_f^2}+\frac{1}{K^2-2(k_0+k\cos\theta)p-m_f^2}
      \right]n(p)\\
    & = & \frac{\alpha}{\pi k}m_fc\int dp(L_1^+(m_f)+L_1^-(m_f))n(p)
  \end{eqnarray*}
  \begin{equation}\end{equation}
  Since the fermions produced in the WIMP annihilation are on-shell, we may use $K^2=m_f^2$
  inside the integrals.  It is then clear from  (\ref{eqL1}) that the integrand vanishes. So,
  we see that at one loop order massless gauge bosons cannot enhance the fermion mass 
  operator. As we will see, all finite coorections will be due to massive particles
  populating the plasma. 
  This brings us to the second term in (\ref{eqhfphoton}) which yields a similar expression 
  that is non zero only for fermions with a non-negligible mass. Thus, we consider only 
  tau and bottom quark production. For the tau
  \begin{equation}
    Re\Sigma_{\beta}^{(2)} = \frac{\alpha}{2\pi k}m_fc\int dp\frac{p}{E}(L_2^+(m_f)+L_2^-(m_f))
    \tilde{n}(E).
  \end{equation}
  where
  \begin{equation}
    L_2^\pm(K,p,E,m_f) \equiv \pm \ln\left(\frac{K^2+m_f^2\pm 2(k_0E+kp)}{K^2+m_f^2\pm 2(k_0E-kp)}
    \right).
  \end{equation}
  and $E=\sqrt{p^2+m_f^2}$. This integral should be evaluated for on-shell fermions with 
  momentum k on the order of the LSP mass.
  In the above expression, the coefficient of $m_fc$ represents the correction to
  the mass operator. This term, too, would vanish identically but for the effect of 
  the non-zero fermion masses. For the tau, this is the full result. The bottom quark 
  receives another, 
  stronger correction from thermal gluon loops.  These will be equal to the above 
  expression with an addional multiplicative 
  factor of $(4/9)(\alpha_s/\alpha)$. For the purpose of this calculations we have 
  considered the LCC1 benchmark point \cite{Weiglein:2004hn} as a cannonical point 
  in the bulk. We have evaluated $\delta m/m$ numerically for 
  the bottom and tau. The results are plotted in Fig.~\ref{figbhflipshift}.  We see that, 
  at freeze-out, the corrections to the fermion mass operators are on the order of $10^{-4}$
  and are thus negligible compared to existing theoretical and experimental uncertainties.
  
  \section{The Width of the A Boson}
  \label{section:AWidth}
  In the so called A-funnel region of mSUGRA parameter space, the mass of the CP-odd heavy 
  Higgs boson ($A^0$) is roughly equal to twice the mass of the LSP and the WIMP annihilation 
  cross section is dominated by resonant s-channel annihilation to $b\bar{b}$, 
  $\tau^+\tau^-$. The value of the 
  cross section is, therefore, highly sensitive to the width of the $A^0$. We would like to investigate 
  the thermal corrections to this quantity and their effect on the relic density predictions.
  
\begin{figure}[t]
    \begin{center}
      \leavevmode
      \epsffile{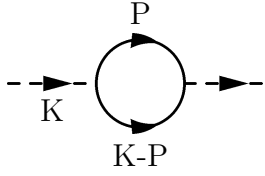}
    \end{center}
    \caption{Feynman diagram contributing to the $A^0$ boson self energy.}
    \label{figfloop}
  \end{figure}

  In the limit that the plasma is populated only by massless gauge bosons and SM 
  fermions, the $A^0$ self energy receives corrections from a single
  Feynman diagram - the fermion loop depicted in Fig.~\ref{figfloop}. We would like to evaluate the 
  imaginary part of this diagram. As discussed in Sec.~\ref{section:ThermalFieldTheory}, 
  we use the finite 
  temperature cutting rules from \cite{Kobes:1986za} to do so. Following the notation 
  in \cite{LeBellac:1996}, we calculate
  \begin{equation}
    \Pi^>  =  -(ie)^2\sum_{f}|G_{fA}|^2\int\frac{d^4P}{(2\pi)^4}Tr\left\{S^>(P)S^>(K-P)\right\}
  \end{equation}
  where the sum is over all the fermion species in the plasma. As usual we consider only
  SM fermions to be thermal. Using (\ref{eqnarraycutprops1}) 
  and (\ref{eqnarraycutprops2}) and some algebra this becomes
  \begin{equation}
    \label{eqAwidth1}
    \Pi^>=-\frac{\alpha}{\pi}|G_{fA}|^2\left(\frac{K^2}{2}-m_f^2\right)
    \frac{1}{k}\int dp\frac{p}{E_1}\left[(1-\tilde{n}_1)(\theta_{-}-\tilde{n}_2)\Theta_{-} +
      \tilde{n}_1(\theta_{+}-\tilde{n}_2)\Theta_{+}\right].
  \end{equation}
  To avoid clutter we have defined 
  \begin{eqnarray*}
    \tilde{n}_{1/2} & = & \tilde{n}(E_{1/2}) \\
    \theta_\pm & = & \theta(k_0\pm E_1)\\
    E_1 & = & \sqrt{p^2+m_f^2}\\
    E_2 & = & \sqrt{(\vec{k}-\vec{p})^2+m_f^2}.
  \end{eqnarray*}
  and we have used the identity
  \begin{equation}
    \epsilon(p_0)(1-\tilde{f}(p_0)) = \theta(p_0)-\tilde{n}(p_0)
  \end{equation}
  The symbols
  $$
  \Theta_\pm = \left\{ 
  \begin{array}{cc}
    1, &\mbox{ if } -1<(K^2\pm2k_0E_1)/2kp<1 \\
    0, &\mbox{ otherwise }
  \end{array}\right.
  $$
  impose kinematic constraints.
  
\begin{figure}[t]
     \begin{center}
     \leavevmode
     \epsffile{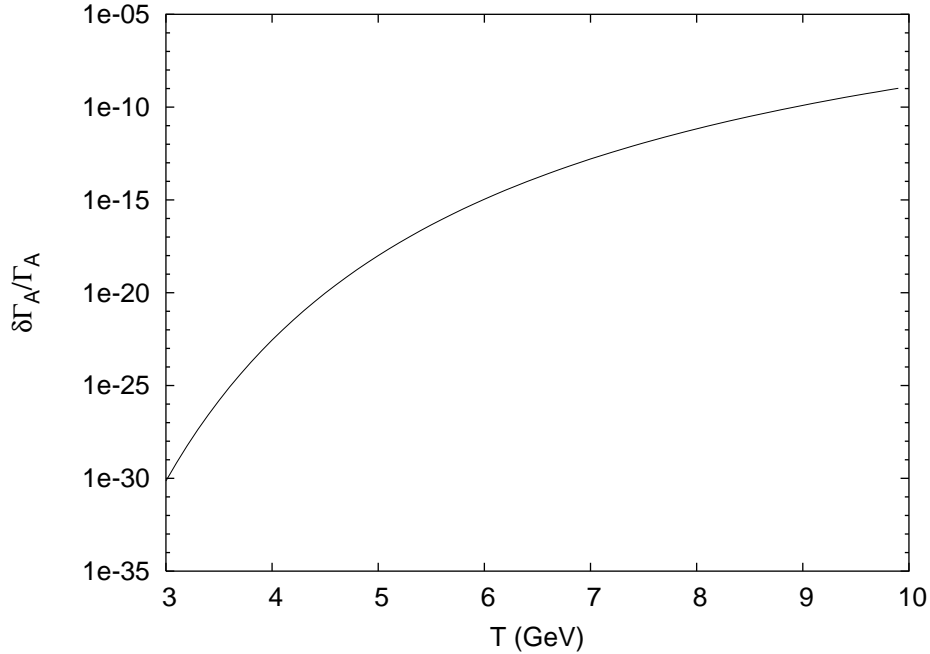}
     \end{center}
     \caption{Thermal correction to the width of the $A^0$ boson.} 
     \label{figdgammaA}
   \end{figure}

   It appears that $\Pi^>$ diverges as $k\to0$. But, in this limit, 
   $\Theta_{+}$ vanishes completely and the region of 
   integration determined by $\Theta_{-}$ collapses to a point given by $K^2=2k_0E_1$. This
   gives a finite result. To evaluate the limit we put the $A^0$ on-shell and set 
   $K=(m,\overrightarrow{0})$. This is a good approximation since $\Gamma_A/m_A$ is 
   typically on the order of 0.05. Writing the infinitesimal region of integration explicitely, 
   we have
   \begin{equation}
     \lim\limits_{k\to0}\Pi^>=\frac{-4\alpha}{\pi}|G_{fA}|^2\left(\frac{m_A^2}{2}-m_f^2\right)
     \frac{1}{k}\int_{q-2k}^{q+2k}dp\frac{p}{E_1}
     (1-\tilde{n}_1)(\theta_{-}-\tilde{n}_2).
   \end{equation}
   Where $q\equiv\sqrt{m_A^2/4-m_f^2}$.  Finally, evaluating the integral, we obtain
   \begin{equation} 
     \lim\limits_{k\to0}\Pi^>=-2\alpha m_A^2\left(1-\frac{2m_f^2}{m_A^2}\right)
     \sqrt{1-\frac{4m_f^2}{m_A^2}}\left(1-\tilde{n}\left(\frac{m_A}{2}\right)\right)^2
     \theta\left(\frac{m_A}{2}-m_f\right).
   \end{equation}
   The imaginary part of the real-time self energy at zero external momentum
   is then \cite{LeBellac:1996}
   \begin{eqnarray*}
     \mbox{Im}\,\bar\Pi & = &-\frac{1}{2}\left(1-e^{-\beta k_0}\right)\Pi^> \\
     & = & \alpha m_A^2|G_{fA}|^2\left(1-\frac{2m_f^2}{m_A^2}\right)
     \sqrt{1-\frac{4m_f^2}{m_A^2}}\left(1-2\tilde{n}_1\left(\frac{m_A}{2}\right)\right)
     \theta(m_A-2m_f).
   \end{eqnarray*}
   \begin{equation} \label{eqAwidth}  \end{equation}
   There are several points worth noting. First, the self energy has no imaginary part for $m_A$ 
   less than $2m_f$.  This is expected for the following reason: at zero external momentum 
   the processes that contribute to the boson's decay width are $A^0\rightarrow f\overline{f}$.
   The additional processes $A^0f\rightarrow \overline{f}$ and $A^0\overline{f}\rightarrow f$,
   corresponding to Landau damping, are kinematically disallowed. Thus, a minimum mass of 
   $m_A>2m_f$ is required for a decay to take place.  Instead of a  power-law dependence 
   on $T$, the thermal $A^0$ width is 
   supressed by a fermionic statistical factor evaluated at half the $A^0$ mass. We could have 
   deduced this fact as well by noting, again, that the only relevant process is the one
   involving the production of a fermion-antifermion pair and its inverse .  These two 
   processes should be accompanied, respectively, by statistical factors 
   $\left(1-\tilde{n}(m_A/2)\right)^2$ and 
   $\tilde{n}(m_A/2)^2$ with a relative minus sign between them.  The total
   statistical factor would then be
   $$
   \left(1-\tilde{n}\left(\frac{m_A}{2}\right)\right)^2-\tilde{n}\left(\frac{m_A}{2}\right)^2 = 
   1-2\tilde{n}\left(\frac{m_A}{2}\right).
   $$
   This is exactly the temperature dependence we obtained in (\ref{eqAwidth}). 
   In Fig.~\ref{figdgammaA} we
   plot the thermal corrections to the A width as a function of temperature. It is clear that at
   freeze-out ($T\approx5$ GeV), the thermal effects are completely negligible.  

   \section{Conclusions}
   \label{section:conclusions}
   In this paper we have identified several mechanisms which lead to heightened sensitivity of the
   WIMP annihilation cross section, and thus the relic density, to the values of the underlying 
   theory parameters. We then investigated the consequences of finite 
   temperature effects in regions of parameter space where these mechanisms are important.  
   We first considered the issue of thermal corrections to parameters to which the relic density 
   is exponentially sensitive, specifically the LSP-NLSP mass splitting in the stau coannihilation 
   region. We found a thermal correction of about $10^{-4}$ to 
   the mass splitting and a similar correction to the relative Boltzmann factor between the stau 
   and the LSP at freeze-out.  This correction is far to small
   to have any measurable consquenses for current or upcoming relic density observations. 
   In the bulk region we considered the issue of corrections to the standard model fermion masses
   and the corresponding shifts in helicity suppressed annihilation cross sections. While
   the branching ratio for these processes is small, the fermions are light and can thus receive 
   significant mass corrections. It is the interplay of these two effects which we investigated.
   We found that, for a typical point in the mSUGRA bulk, the thermal effects correct the 
   fermionic mass operators by at most a factor of $10^{-4}$ at freeze-out. Finally, we 
   considered thermal corrections to the width of the $A^0$ boson in the A-funnel region.
   We found the value of this parameter to be exponentially suppressed by the ratio of the 
   A mass to the freeze-out temperature and we provided a physical interpretation for this result.

   Although in the specific family of models which we studied all thermal effects are considerably below
   the detection threshold of any current or planned observation of the relic density, we believe that
   generically these
   issues should be considered.  For example, in any coannihilation scenario an exponential dependence 
   on a small mass splitting is expected.  Similarly, any resonant annihilation will exhibit hightened
   sensitivity to a small width associated with the virtual particle. Small corrections to these
   parameters cannot a priori be neglected.  However, as we have seen, close analysis will
   often indicate that these effects are very small.
   
   \section{Acknowledgments}
   I would like to thank Edward Baltz for explaining the subtleties of DarkSUSY to me, 
   Renata Kallosh for enlightening discussions and Michael Peskin for many helpful comments
   and suggestions. This work was supported by the US Department of Energy,
   contract DE--AC02--76SF00515.

\end{document}